\documentclass[twocolumn]{aastex631}
\usepackage{soul, color}
\soulregister\citet7
\soulregister\citep7
\soulregister\ref7
\submitjournal{ApJ}

\graphicspath{{./}{figures/}}
\shorttitle{DFUWS Galaxies}
\shortauthors{Shen et al.}
\begin{document}
\title{First Results from the Dragonfly Ultrawide Survey: the Largest Eleven Quenched Diffuse Dwarf Galaxies in 3100 deg$^2$ with Spectroscopic Confirmation}

\correspondingauthor{Zili Shen}
\email{zili.shen@yale.edu}

\author[0000-0002-5120-1684]{Zili Shen}
\affiliation{Department of Astronomy, Yale University, 
New Haven, CT 06520, USA}

\author[0000-0003-4381-5245]{William P. Bowman}
\affiliation{Department of Astronomy, Yale University, New Haven, CT 06520, USA}

\author[0000-0002-8282-9888]{Pieter van Dokkum}
\affiliation{Department of Astronomy, Yale University, New Haven, CT 06520, USA}

\author[0000-0002-4542-921X]{Roberto Abraham}
\affiliation{David A. Dunlap Department of Astronomy \& Astrophysics, University of Toronto, Toronto, ON M5S3H4, Canada}
\affiliation{Dunlap Institute, University of Toronto, 50 St. George Street, Toronto, ON M5S3H4, Canada}

\author[0000-0002-7075-9931]{Imad Pasha}
\affiliation{Department of Astronomy, Yale University, New Haven, CT 06520, USA}
\author[0000-0002-7743-2501]{Michael A. Keim}
\affiliation{Department of Astronomy, Yale University, New Haven, CT 06520, USA}

\author[0000-0002-7490-5991]{Qing Liu}
\affiliation{David A. Dunlap Department of Astronomy \& Astrophysics, University of Toronto, Toronto, ON M5S3H4, Canada}
\affiliation{Dunlap Institute, University of Toronto, 50 St. George Street, Toronto, ON M5S3H4, Canada}

\author[0000-0002-2406-7344]{Deborah M. Lokhorst}
\affiliation{NRC Herzberg Astronomy \& Astrophysics Research Centre, 5071 West Saanich Road, Victoria, BC V9E2E7, Canada}

\author[0000-0003-0327-3322]{Steven R. Janssens}
\affiliation{Centre for Astrophysics \& Supercomputing, Swinburne University, Hawthorn, VIC 3122, Australia} 
\affiliation{ARC Centre of Excellence for All Sky Astrophysics in 3 Dimensions (ASTRO 3D), Australia}

\author[0000-0002-4175-3047]{Seery Chen}
\affiliation{David A. Dunlap Department of Astronomy \& Astrophysics, University of Toronto, Toronto, ON M5S3H4, Canada}
\affiliation{Dunlap Institute, University of Toronto, 50 St. George Street, Toronto, ON M5S3H4, Canada}

\begin{abstract}

The Dragonfly Telephoto Array employs a unique design to detect very large and diffuse galaxies, which might be missed with conventional telescopes. 
%Limitations in sky subtraction and the presence of scattered light and chip boundaries in existing surveys make it difficult to look for objects that are larger than $\sim30\arcsec$ in diameter. 
The Dragonfly Ultrawide Survey (DFUWS) is a new wide-field survey which will cover 10,000 deg$^2$ of the northern sky, and it provides an ideal dataset to find these large diffuse galaxies. 
From 3100 deg$^2$ of DFUWS data, we identified eleven large, low surface brightness galaxies as a pilot sample for spectroscopic follow-up. These are the largest galaxies in the examined area that appear smooth and isolated, with effective radii of 12\arcsec-27\arcsec. Eight are below 24 $\mathrm{mag\,arcsec^{-2}}$ in central $g$-band surface brightness.
Keck Cosmic Web Imager (KCWI) spectra of the diffuse light show that all eleven galaxies in this sample are quiescent, and seven qualify as ultra-diffuse galaxies (UDGs). 
Eight galaxies have distances between 15 and 30\,Mpc, while the other three are in the Pegasus cluster at 50\,Mpc.
Their spectra show evidence of a $\sim 1$\,Gyr old stellar population in addition to an even older stellar population. The intermediate-age component is present in group and satellite galaxies but not in the Pegasus cluster UDGs.
All galaxies in this sample are detected in both Dragonfly and Legacy imaging, and the sample partially overlaps with existing UDG catalogs. 
This pilot sample provides an excellent training set for our analysis of the upcoming full 10,000 deg$^2$ DFUWS data, from which we may expect to discover even larger, previously-unknown galaxies.  

\end{abstract}


\section{Introduction} \label{sec:intro}
Large, low surface brightness (LSB) galaxies remain elusive in optical surveys despite being a significant component of the galaxy population. The central surface brightness of these galaxies are $\sim 100$ times fainter than the night sky, and special optical design and data reduction pipelines are needed for LSB imaging, especially to detect LSB galaxies with large apparent sizes. The two most comprehensive catalogs of LSB galaxies \citep{Zaritsky2023ApJS, Tanoglidis2021ApJS} become incomplete at angular sizes $R_{\mathrm{eff}}> 15\arcsec$. 
%However, these large and faint galaxies are extremely valuable for understanding the formation and evolution of LSB galaxies: one key question is whether they require novel (and potentially exotic) formation pathways or if they represent the extreme end of known phenomena.
%A fundamental challenge in studying the intrinsic properties of LSB galaxies is to determine the distance.
An LSB galaxy with apparent size $>15\arcsec$ is either a dwarf galaxy at $\sim 5$\,Mpc, a more distant ($15\sim30$\,Mpc) ultra diffuse galaxy (UDG), or a true giant at $\sim 100$\,Mpc, such as Malin 1 \citep{Bothun1987AJ}. 
Since the \citet{vanDokkum2015ApJL} study on Coma UDGs, there has been a renewed interest in this population of physically large ($Re>1.5$kpc), LSB ($\mu_0,g>24$ mag arcsec) galaxies. Most known UDGs are at distances of $50-100$\,Mpc, and concentrated in galaxy clusters. Follow-up studies that constrain the underlying dark matter halo properties (e.g. velocity dispersion) or the star formation history (e.g. stellar population analysis) are difficult at that distance. Even at 20 Mpc, resolved studies of individual UDGs \citep{Danieli2019ApJL,Shen2023ApJ} require long integration times on the world's largest telescopes. Measuring the distance of the LSB galaxies requires deep spectra or at least semi-resolved stellar populations, both of which are observationally expensive. 

\begin{figure*}
    \centering
    \includegraphics[width=\textwidth]{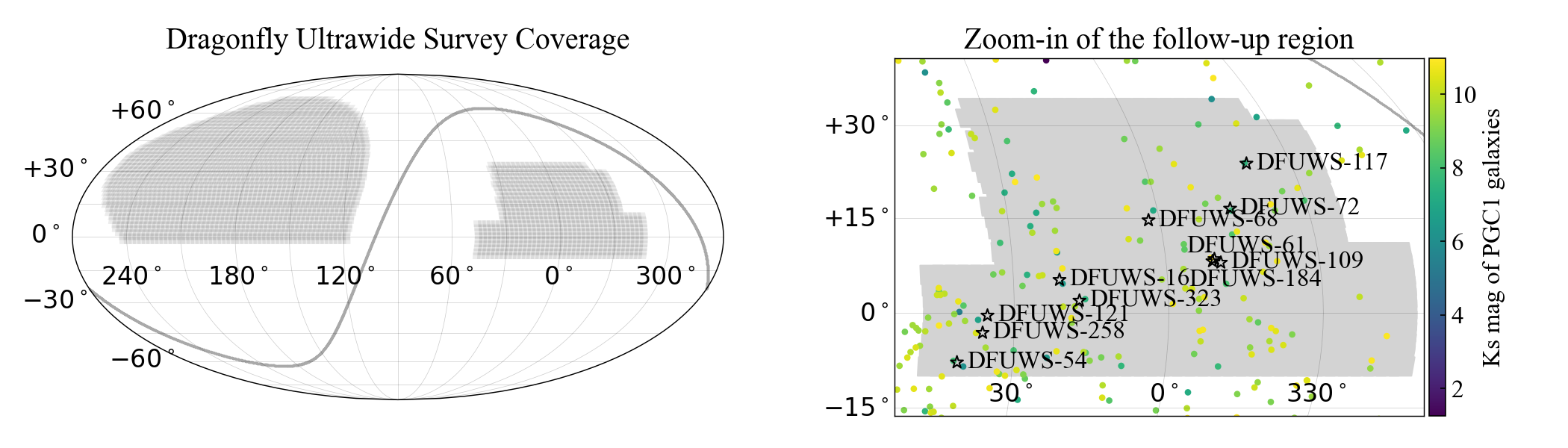}
    \caption{Left panel: Dragonfly Ultrawide survey coverage on sky (RA and Dec). Each small grey rectangle within the footprint shows the size of one Dragonfly pointing to scale. The solid grey line marks the Galactic plane. Right panel: a zoomed-in view of the Survey footprint with black stars showing the location of galaxies presented in this paper. Each galaxy is labelled with its ID number. Colored dots are known groups from the PGC1 catalog \citep{Kourkchi2017ApJ} with $K_s < 11$ mag.}
    \label{fig:survey-map}
\end{figure*}

To probe the formation of LSB galaxies, we need a relatively nearby sample. To find them in optical imaging without \textit{a priori} knowledge of distances, our strategy is to look for LSB galaxies with the biggest \textit{apparent} sizes.
%For galaxies of a fixed type (e.g. UDGs), the largest among them is the closest in distance; for galaxies at a fixed distance, the galaxy with the largest angular size is intrinsically the biggest.
%Thus, selecting objects with the largest apparent size yields a sample that can be followed up easily. 
However, there is an upper limit on the biggest galaxy that can be detected in existing optical surveys. 
LSB imaging requires careful control of systematics and sky subtraction, which is difficult to maintain over a wide area. Due to the telescope field of view and choices in the data analysis, LSB galaxy surveys \citep{Zaritsky2023ApJS, Tanoglidis2021ApJS} focus on galaxies with angular sizes $R_{\mathrm{eff}}< 15\arcsec$ and they do not detect any galaxies larger than $30\arcsec$.

The Dragonfly Telephoto Array enables the extension of the LSB surveys to larger galaxy sizes. Small robotic telescopes have been productive in the LSB regime, from finding dwarf galaxies \citep[DGSAT,][]{Javanmardi2016A&A} to studying the haloes and environments of nearby galaxies \citep[HERON,][]{Rich2019MNRAS} to extragalactic stellar streams \citep[Condor,][]{Lanzetta2024MNRAS}. Dragonfly led to the discovery of UDGs in the Coma cluster \citep{vanDokkum2015ApJL} and dwarf galaxies in nearby groups 
\citep{Merritt2014ApJL, Danieli2017ApJ, Cohen2018ApJ}.  Dragonfly was designed to reach surface brightness levels below $\sim 28-29$ mag arcsec$^{-2}$ \citep{Abraham2014PASP}. 
The high-end telephoto lenses at the heart of its design and the unobstructed optical path minimize artifacts in Dragonfly data. Thanks to the wide field-of-view ($3^{\circ}\times2^{\circ}$), structures on scales up to $\sim 45\arcmin$ are preserved in the data reduction process \citep[see][]{Danieli2020ApJ}. LSB galaxies that are arcminute-sized are comfortably within this range, allowing us to extend the LSB search into unexplored territory.

In addition to searching previously overlooked surface brightness vs. size parameter space, it is also important to conduct searches over a wide area on sky. Environment is critical to LSB galaxy formation and evolution. Tidal interactions with a massive galaxy or with a galaxy cluster are thought to turn normal dwarf galaxies into UDGs \citep{Conselice2018RNAAS,Carleton2019MNRAS}. Field UDGs have been identified both in optical imaging and in HI, and are bluer than UDGs in dense environments \citep{Prole2019MNRAS, Kadowaki2021ApJ}. Quenched field UDGs are extremely rare and have recently been proposed to be splash-back galaxies that passed through a dense environment \citep{Benavides2021NatAs}. This demonstrates that dense environments are not necessary for the formation of large LSB galaxies, but they may be required for quenching the star formation in these galaxies.

In this paper, we report our initial efforts to find a nearby sample of large LSB galaxies across a range of environments. To accomplish this, we must first identify a sample obtained over a large enough sky area to capture all environments. We conduct the search using the Dragonfly Ultrawide Survey (DFUWS), which maintains LSB sensitivity over a wide area (Section \ref{sec:phot}). For this paper, we report on early results from one third of the DFUWS footprint (the full sample will be described in a future paper). Having obtained this sample, our focus then turns to reliably measuring the distance to the galaxies. We report radial velocities from Keck/KCWI spectra for the entire sample (Section \ref{sec:spec}). Distances and environments of each galaxy are presented in Section \ref{sec:results}. Finally, Section \ref{sec:dis} discusses the implications from our analysis for the abundance of large quiescent UDGs in the local Universe, along with evidence for multiple stellar populations in this sample.
%Throughout this paper, we assume a concordant cosmological model with $H_0=75$, $\Omega_\mathrm{m}=0.27$,$\Omega_\mathrm{\Lambda}=0.73$ (from Cf3)

\section{Survey and Photometric Analysis} \label{sec:phot}

\subsection{The Dragonfly Ultrawide Survey}

% write about Ultrawide
The new galaxies in this paper were discovered in DFUWS, which will be described in detail in an upcoming paper (Bowman et al., in prep). We will give a brief overview of the data here. 
The Dragonfly Telephoto Array \citep{Abraham2014PASP} is composed of 48 Canon 400\,mm\,$f/$2.8 ISII USM-L telephoto lenses, each attached to an SBIG KAF-8300 CCD camera.
Half of these lenses are equipped with a SDSS $g$ filter and the other half with a SDSS $r$ filter. 
Each lens and camera exposes simultaneously with small pointing offsets, so that the entire array has a field of view of $2\arcdeg \times 3\arcdeg$ and a pixel scale of $2.\arcsec5 \,\mathrm{pix}^{-1}$ after resampling.
The lack of reflective surfaces inside the lenses, their unobstructed light paths, and their nanofabricated coatings that suppress internal reflection all contribute to a well-behaved wide angle point-spread-function (PSF) \citep{Liu2022ApJ}. The wide angle PSF model is critical to subtracting high surface brightness compact sources so that faint structures can be detected.

%Observing is done semi-automatically each night, where the observer manually runs an observing script that takes calibration frames, monitors observing conditions, and chooses targets. At the end of each night, a script automatically reduces the new raw frames and generates coadded images for quality control. 

Dragonfly data are reduced using an updated version of the \texttt{DFReduce} pipeline \citep{Danieli2020ApJ}.
Data from each camera are reduced independently until the final coadded image stack is made. 
The pipeline handles sky subtraction in two iterations to ensure that all structures below 45$\arcsec$ are preserved. 
While the \texttt{DFReduce} upgrades since \citet{Danieli2020ApJ} will be described in Bowman et al. (in prep), we summarize the major updates here. 
The pipeline now accommodates parallel reduction runs so that the entire survey can be reduced together.
All of the raw data are stored on Amazon Web Service (AWS) and the reduction is run using AWS Batch processing and orchestrated by AWS database servers.
The reduced coadds are scaled to a uniform zeropoint and tiled into $1\arcdeg \times 1\arcdeg$ mosaics. These tiles are the final data product of DFUWS and the input data in our target selection process.

DFUWS is designed to cover the entire SDSS imaging footprint, shown in the left panel of Figure \ref{fig:survey-map}. The integration time for each pointing is 30 minutes split into three 600s exposures. In theory, this yields 144 total frames (72 for each band), and the minimum requirement for a field is 50 frames for each band to account for nonoperational units and rejected frames. The observation campaign started in 2019 and the full data release is projected for Jan 2025. 
The reduced data reach a $1\sigma$ surface brightness depth of $29.5 \,\mathrm{mag\,arcsec^{-2}}$ on arcmin ($1\arcmin \times 1 \arcmin$) scales in the $g$-band\footnote{The depth is defined as measured fluctuations in an image at a particular scale, and was implemented by \citet{Keim2022ApJ}.}.
This paper focuses on the 3100 deg$^2$ area around the South Galactic Cap, shown in the right panel of Figure \ref{fig:survey-map}. This area is the early science data that had been observed and reduced by Oct 2023, allowing us to find the first batch of follow-up targets.

\subsection{Target selection}
\begin{figure*}[ht!]
    \centering
    \includegraphics[width=\textwidth]{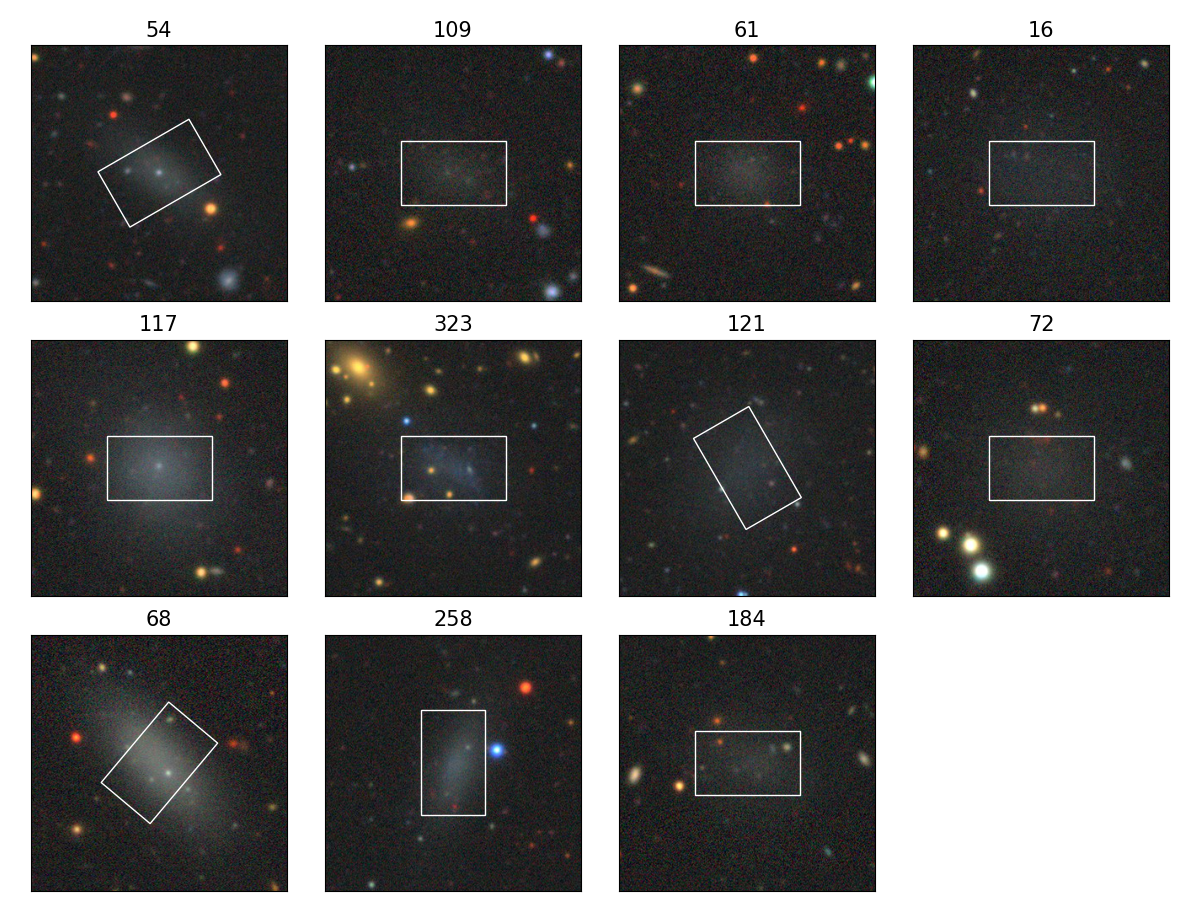}
    \caption{Legacy cutouts of all DFUWS galaxies in this paper. The white rectangles represent the KCWI field of view and orientation during our Oct 2023 observing run.}
    \label{fig:Legacy}
\end{figure*}

In imaging surveys, low-surface-brightness galaxies are typically detected by visual inspection, automatic detection, or a combination of both.
The key challenge to a fully automated detection algorithm is that widely-used source detection methods can shred a low surface brightness galaxy into multiple components and assign parts of it to nearby high surface brightness point sources.
Automated algorithms are also affected by image artifacts that can lead to a high false-positive rate \citep{Zaritsky2023ApJS}. 
To remove the false-positive detections, the catalogue needs to be cleaned using either preset cuts on galaxy properties, machine learning, or visual inspection of the detected objects. 
Direct visual inspection of images has the advantage of requiring no prior knowledge of the galaxy properties and generally results in a higher purity of the catalogue. The downside is that sample completeness, and its dependence on parameters like surface brightness, is hard to characterize. 

Since this is the first analysis of data from DFUWS, we visually inspected $r$-band images of the 3100 deg$^2$ area shown in the right panel of Figure \ref{fig:survey-map}.
Two factors made visual inspection feasible for an area this large.
Firstly, faint galaxies stand out in Dragonfly images as fuzzy blobs. Dragonfly images are very clean, with no diffraction spikes or artifacts. Under the right display scaling, LSB features appear ``fuzzy'' (blending into the sky) while high surface brightness sources have a solid edge. 
Secondly, the large pixel scale of Dragonfly means that each image covers a large area on sky. In practice, we found that binning the images $2\times2$ did not affect the visual detection results. We inspected $1\arcdeg \times 1\arcdeg$ images which were only $750\times750$ pixels. This allowed for very fast image display at standard data transfer speeds.

Multiple team members were involved in the visual inspection and each person was trained by inspecting tiles with known UDGs in them \citep[e.g. the NGC1052 field,][]{Cohen2018ApJ}.
The team members first familiarized themselves with the images and learned to distinguish LSB galaxies from galactic cirrus. 
After the training period, we adopted a two-step process for LSB galaxy detection. The first step was adjusting the display scaling until LSB features appeared distinctively fuzzy.
Due to the low spatial resolution of our data, these LSB features could be unresolved background spiral galaxies, high redshift galaxy clusters, or galactic cirrus blended with a star.
Thus, the second step was to verify a potential LSB galaxy candidate by viewing the image at the corresponding coordinates from the Legacy Survey \citep{Dey2019AJ}. 
Each tile was randomly assigned to a team member and each object found was compiled into a combined catalog.

To ensure the uniformity of the sample and to remove any additional contaminants, the lead author reviewed the combined catalogue from all team members. After removing duplicates from overlapping tiles, the final visual catalogue contained 314 LSB galaxy candidates. We then cross-matched the catalogue to SIMBAD and obtained prior photometric information where available. For the candidates with no prior record in SIMBAD, we visually estimated their size and surface brightness and rejected candidates that were clumpy, too bright, or too small, leaving 56 candidates. We downloaded Legacy $g$-band images for these candidates and fit a preliminary S{\'e}rsic model. From these candidates, we assigned them a priority for obtaining Keck spectroscopic follow-up. ``High priority'' targets had no prior radial velocity measurement, preliminary surface brightness below 24\,$\mathrm{mag\,arcsec^{-2}}$, sizes greater than 15$\arcsec$, and no massive galaxy neighbors. ``Low priority'' targets were either brighter or smaller but appeared isolated. ``Group'' candidates were close to a known galaxy group but had not been identified, or spectroscopically observed. We note that any categorization at this point was preliminary due to the lack of radial velocity and distance estimates.
Due to practical constraints (RA distribution, varying exposure time per target), the final spectroscopically observed sample of eleven objects contains all the high priority targets and several from other categories. 

\subsection{Photometry}
\begin{figure*}
    \centering
    \includegraphics[width=\textwidth]{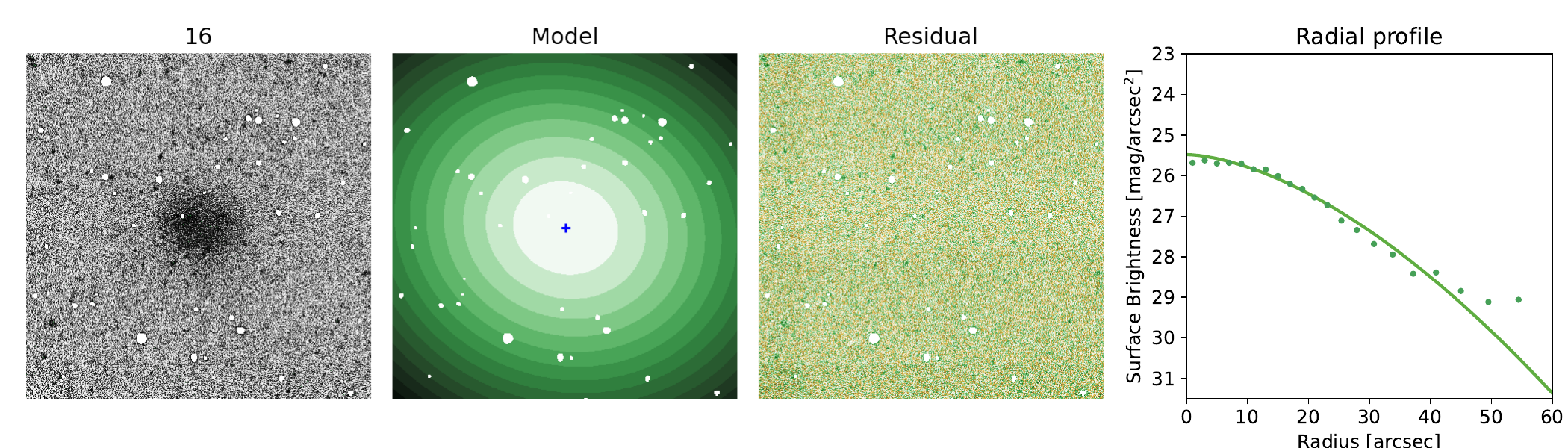}
    \caption{Legacy $g$-band image and model of DFUWS-16. The complete figure set (11 images) is available in the online journal.}
    \label{fig:phot}
\end{figure*}

The eleven galaxies in our spectroscopically observed sample are all detected in both Dragonfly and Legacy.  
The left panel in Figure \ref{fig:phot} shows a Legacy $g$-band cutout centered on each galaxy, as well as a S{\'e}rsic model that fit the image.
We chose to report the photometric measurements from Legacy imaging because of its higher spatial resolution. Many of the galaxies have star clusters (54, 117, 68) or nearby contaminants (323, 258, 54) which would be blended with the diffuse light of the galaxy in the Dragonfly image.
The Dragonfly cutouts are shown in the left column of Figure \ref{fig:spec} and we discuss a technique to remove the contaminants in Section \ref{subsec:mrf}.

We used the \texttt{AstroPhot} package \citep{Stone2023MNRAS} to find the best-fit S{\'e}rsic model and to extract the light profile, shown in Figure \ref{fig:phot}.
Before fitting, we used \texttt{sep} to detect sources in the image and mask all other sources. Each galaxy is fit with a single S{\'e}rsic model unless it needed a second component.
DFUWS-54 and DFUWS-117 had nuclear star clusters which were separately modeled from the diffuse light of the galaxy.
For those cases, we added a point source component based on the PSF from the Legacy Survey at the location of each galaxy. The residual plot shows the data subtracted by the full model, while the profile in Figure \ref{fig:phot} shows the S{\'e}rsic model. 
DFUWS-323 was spatially close to an elliptical galaxy which meant that a joint model was needed. We added a second S{\'e}rsic model for the elliptical so that the fit to DFUWS-117 would not be biased by the outskirts of the elliptical galaxy.

The resulting photometric measurements are presented in Table \ref{table:phot}. All measurements except the $g-r$ color are solely based on the fit to the $g$-band Legacy data. 
Galaxies in this sample have half-light radii mostly between 15$\arcsec$ and 22.5$\arcsec$, with one galaxy that is smaller (DFUWS-61 at 12.3\arcsec) and one galaxy that is bigger (DFUWS-72 at 27.7\arcsec) than the rest.
This sample also has consistently red colors ($g-r>0.5$), with DFUWS-323 being the bluest ($g-r=0.24$).
All galaxies in our sample have half-light radii smaller than $30\arcsec$ and we do not find any arcminute-sized galaxies in these 3100 deg$^2$.

We cross-matched our sample with the dwarf galaxy catalog from \citet{Paudel2023ApJS}, who conducted a systematic visual inspection of Legacy Survey data.
This sample also partially overlaps with the SMUDGes catalog \citep{Zaritsky2023ApJS}.
These external catalog were not taken into account during our selection procedure. After the eleven spectroscopic follow-up galaxies were selected, we found a match for all galaxies except 109, 61, and 184.
The smooth and featureless morphology of these galaxies suggest that they are gas-poor and quenched. 
To establish the physical size and stellar populations of these galaxies, we turned to spectroscopic folllow-up, as described next.

\begin{deluxetable*}{cccccccccc}
\tablecaption{\label{table:phot} Photometric properties of the biggest DFUWS galaxies in the South Galactic Cap region. The magnitudes are measured from Legacy data. [PYY2023] designates objects in \citet{Paudel2023ApJS} and SMDG refers to the SMUDGes catalog \citep{Zaritsky2023ApJS}. PegI-UDG03 was first identified in \citet{Shi2017ApJ}.}
\tablehead{\colhead{ID} & \colhead{Name} & \colhead{R.A.} & \colhead{Decl.} & \colhead{$m_g$} & \colhead{$\mu_0 (g)$} & \colhead{$g-r$} & \colhead{$R_e$} & \colhead{Sersic $n$} & \colhead{Ellipticity}\\ \colhead{ } & \colhead{ } & \colhead{$\mathrm{J2000}$} & \colhead{$\mathrm{J2000}$} & \colhead{$\mathrm{mag}$} & \colhead{$\mathrm{mag\,arcsec^{-2}}$} & \colhead{$\mathrm{mag}$} & \colhead{$\mathrm{arcsec}$} & \colhead{ } & \colhead{ }}
\startdata
68 & [PYY2023] J000844+143529 & 00$^{\rm h}08^{\rm m}44.56^{\rm s}$ & +14\arcdeg35\arcmin30\farcs45 & 16.67 & 23.46 & 0.68 & 18.0 & 0.67 & 0.44 \\
323 & SMDG J0109171+015300 & 01$^{\rm h}09^{\rm m}17.09^{\rm s}$ & +01\arcdeg52\arcmin59\farcs22 & 17.41 & 24.42 & 0.24 & 18.5 & 0.88 & 0.70 \\
16 & [PYY2023] J012412+050910 & 01$^{\rm h}24^{\rm m}13.00^{\rm s}$ & +05\arcdeg09\arcmin10\farcs86 & 17.62 & 25.48 & 0.55 & 20.2 & 0.61 & 0.86 \\
121 & SMDG J0220482-002754 & 02$^{\rm h}20^{\rm m}48.20^{\rm s}$ & -00\arcdeg27\arcmin54\farcs17 & 17.40 & 24.94 & 0.54 & 21.8 & 0.87 & 0.80 \\
258 & SMDG J0224231-031100 & 02$^{\rm h}24^{\rm m}23.18^{\rm s}$ & -03\arcdeg10\arcmin58\farcs55 & 17.76 & 23.83 & 0.54 & 16.5 & 1.00 & 0.45 \\
54 & [PYY2023] J024321-075032 & 02$^{\rm h}43^{\rm m}21.86^{\rm s}$ & -07\arcdeg50\arcmin32\farcs54 & 17.80 & 24.00 & 0.58 & 18.4 & 0.99 & 0.42 \\
117 & LEDA   87226 & 22$^{\rm h}36^{\rm m}11.74^{\rm s}$ & +23\arcdeg42\arcmin42\farcs36 & 16.44 & 23.56 & 0.57 & 18.6 & 1.06 & 0.80 \\
72 & MATLAS-2176 & 23$^{\rm h}01^{\rm m}45.12^{\rm s}$ & +16\arcdeg27\arcmin22\farcs92 & 17.25 & 25.05 & 0.65 & 27.7 & 0.92 & 0.97 \\
109 &  & 23$^{\rm h}17^{\rm m}15.56^{\rm s}$ & +07\arcdeg52\arcmin16\farcs95 & 18.38 & 25.14 & 0.67 & 19.0 & 0.98 & 0.60 \\
61 & PegI-UDG03 & 23$^{\rm h}21^{\rm m}57.39^{\rm s}$ & +08\arcdeg26\arcmin27\farcs33 & 18.45 & 24.82 & 0.61 & 12.3 & 0.79 & 0.78 \\
184 &  & 23$^{\rm h}23^{\rm m}40.29^{\rm s}$ & +08\arcdeg05\arcmin43\farcs40 & 17.94 & 25.16 & 0.71 & 19.7 & 0.85 & 0.71
\enddata
\end{deluxetable*}

\section{Spectroscopic Data and Analysis} \label{sec:spec}
\subsection{Observation}
LSB galaxy spectroscopy has long been a challenge \citep{Kadowaki2021ApJ}, but the Keck Cosmic Web Imager KCWI provides a highly sensitive ``light bucket'' mode. Despite the LSB nature of these objects, we can build up signal-to-noise by collecting galaxy light within the KCWI field-of-view (FOV) and then averaging over all pixels \citep[e.g.][]{Gannon2023MNRAS, Shen2023ApJ}.
Spectra of all eleven galaxies in the sample were obtained with KCWI on Keck II in October 2023.
The large image slicer was used with the low-resolution BL grating, resulting in a FOV of $33\arcsec \times 20\arcsec$
and an approximate spectral resolution of $R\sim900$. 
The central wavelength was 4500\AA\ and the wavelength coverage was $3500-5500$\AA.
Data were taken with $2\times2$ binning and the sky position angles are shown in Fig. \ref{fig:Legacy}. The non-zero position angles were chosen to align with the minor axis of the galaxy, originally for better sky subtraction, though in practice we found that sky offset exposures were still necessary, because many galaxies in our sample were bigger than the FOV. We therefore obtained offset sky pointings for each galaxy. The offset sky positions were chosen to be empty of stars and galaxies and at least $3\times R_{e}$ from the center of the galaxy. Each science pointing was sandwiched between two offset sky pointings, and the average of these two adjacent sky spectra were used for sky subtraction. The exposure time for each science and sky pointing is 600\,s. UDG-16 and UDG-72 both had four science exposures, while all other galaxies had one or two science exposures. The observing conditions were excellent, with clear skies on both nights when these data were taken.

\subsection{Radial velocities}
\label{sec:rv}
\begin{figure*}
    \centering
    \includegraphics[width=0.95\textwidth]{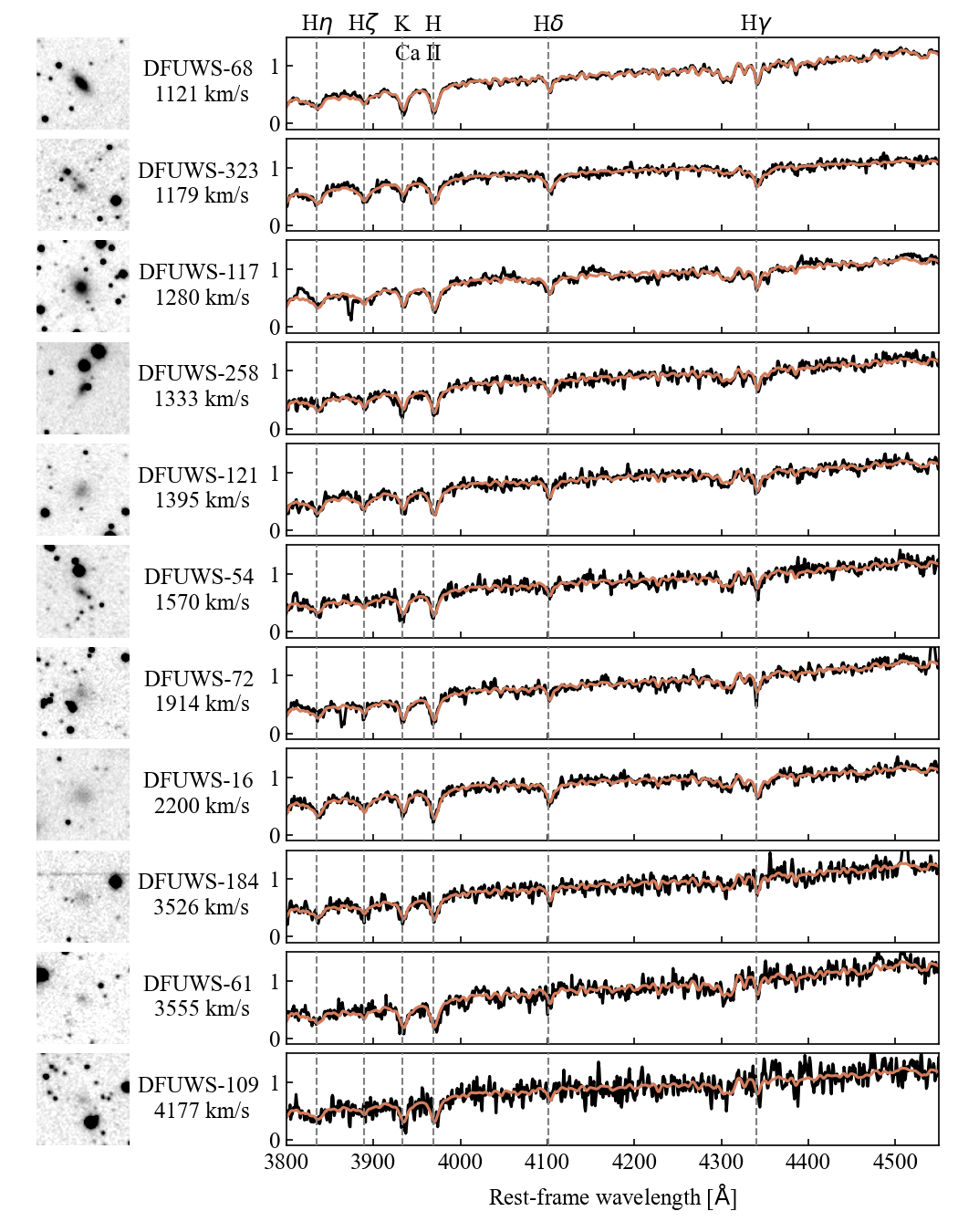}
    \caption{Left column: Dragonfly g-band cutouts of each galaxy in our sample. Right column: Keck/KCWI spectra of each galaxy (black line) and the best-fit pPXF model (orange line). Both the data and the model have been de-redshifted to the rest frame, using the radial velocity listed in each row. We label Balmer lines and the Ca\,II H\&K lines to guide the eye.}
    \label{fig:spec}
\end{figure*}
Each datacube was reduced and calibrated using the KCWI Data Reduction Pipeline (KCWI DRP)
\footnote{\url{https://github.com/Keck-DataReductionPipelines/KCWI_DRP}} in default mode with the option to skip sky subtraction turned on.
The KCWI DRP reduces each of the science and sky frames independently, using 
``Bars'' exposures to map each pixel in the raw image into position in the data cube and the arc lamp exposures to find the wavelength solution.
These transformations were used to convert the 2D science image into a data cube with three dimensions: slice number, position along the slice, and wavelength. 

These data cubes, dubbed \texttt{icubed} files by the KCWI DRP were then used for sky subtraction. 
Sky subtraction is one of the most important steps in spectroscopy of LSB sources. We found that using sky pixels in the edge of the field did not work well, either due to the extended size of these galaxies or systematics with the KCWI detector. 
Although the offset sky exposures added overhead for the observing time, we found it to be the cleanest method of sky subtraction.
After rejecting $5\sigma$ outlier pixel values at each wavelength, each data cube was averaged into a 1D spectrum.
The average sky spectrum from two adjacent offset exposures was subtracted from each science spectrum. For galaxies with multiple science exposures, all sky-subtracted science spectra were averaged. The final spectrum for each galaxy is shown as the black line in the right column of Fig. \ref{fig:spec}.
The most prominent absorption features in these spectra are the Balmer absorption lines, the Ca\,II H\&K lines just blueward of rest-frame 4000\AA, and the G-band at about 4300\AA.

The radial velocity was obtained by fitting the final spectra of each galaxy with \texttt{pPXF} \citep{Cappellari2023MNRAS}. \texttt{pPXF} performs full-spectral fitting by convolving simple stellar population (SSP) templates with a range of line-of-sight velocities ($v$) and velocity dispersions ($\sigma$) to produce the best fit to the data. We used the template spectra of \texttt{FSPS} \citep{Conroy2009ApJ, Conroy2010ApJ}, which in the spectral range of our data ($3500$\AA\ to $5500$\AA) are based on the MILES stellar library \citep{Sanchez-Blazquez2006MNRAS}. We limited the metallicity range of the templates to $-1.5<[M/H]<0.5$ and left the age range to the full 1 to 17.78 Gyr. A fourth-order multiplicative polynomial was allowed in the fit to account for the continuum.

The best-fit model is shown as the orange line in Fig. \ref{fig:spec}. The heliocentric radial velocities are listed in each row of Fig. \ref{fig:spec} and in Table \ref{table:dist}. The uncertainties in the radial velocity were obtained from bootstrap analysis. We randomly added back residuals from the fiducial fit and re-fit for 30 iterations, and we report the standard deviation of the resulting velocities as the error. 
We also report in Table \ref{table:dist} the distance calculated with Cosmicflows-3 \citep{Kourkchi2020AJ} from the radial velocities. 
The uncertainty in the radial velocities translates to $\sim0.1$\,Mpc in distance, but a bigger source of uncertainty is the unknown peculiar velocity with respect to the Cosmicflows-3 model, where each 100$\,\mathrm{km\,s^{-1}}$ results in $1-2$\,Mpc error in distance.
From the fiducial value, six out of eleven galaxies are within 20\,Mpc, eight are within 30\,Mpc, while the farthest three are at $40-50$\,Mpc. 

\section{Results} \label{sec:results}
\subsection{Classification}
\label{subsec:class}

\begin{deluxetable*}{ccccccccc}
\tablecaption{\label{table:dist} Photometric properties of the biggest DFUWS galaxies in the South Galactic Cap region. The magnitudes are measured from Legacy data. [PYY2023] designates objects in \citet{Paudel2023ApJS} and SMDG refers to the SMUDGes catalog \citep{Zaritsky2023ApJS}. PegI-UDG03 was first identified in \citet{Shi2017ApJ}.}
\tablehead{\colhead{ID} & \colhead{RV} & \colhead{[M/H]} & \colhead{Distance} & \colhead{$R_e$} & \colhead{$M_g$} & \colhead{Classification} & \colhead{Environment}\\ \colhead{ } & \colhead{$\mathrm{km\,s^{-1}}$} & \colhead{ } & \colhead{$\mathrm{Mpc}$} & \colhead{$\mathrm{kpc}$} & \colhead{$\mathrm{mag}$} & \colhead{ } & \colhead{ }}
\startdata
68 & 1121$\,\pm\,$7 & -0.88 & 14.9 & 1.30 & -14.19 & Dwarf galaxy & NGC\,7814 \\
323 & 1179$\,\pm\,$16 & -1.19 & 15.1 & 1.36 & -13.49 & LSB dwarf & Field \\
258 & 1333$\,\pm\,$13 & -1.11 & 15.4 & 1.23 & -13.18 & Dwarf galaxy & Field \\
121 & 1395$\,\pm\,$9 & -1.03 & 16.2 & 1.71 & -13.64 & UDG & Field \\
117 & 1280$\,\pm\,$6 & -0.98 & 17.5 & 1.58 & -14.78 & Dwarf galaxy & NGC\,7332 \\
54 & 1570$\,\pm\,$14 & -1.07 & 18.3 & 1.63 & -13.51 & UDG & NGC\,988 \\
72 & 1914$\,\pm\,$9 & -1.07 & 24.2 & 3.25 & -14.67 & UDG & NGC\,7454 \\
16 & 2200$\,\pm\,$7 & -0.92 & 26.9 & 2.64 & -14.53 & UDG & NGC\,488 \\
184 & 3526$\,\pm\,$8 & -1.06 & 44.3 & 4.23 & -15.29 & UDG & Pegasus \\
61 & 3555$\,\pm\,$11 & -1.04 & 44.7 & 2.66 & -14.81 & UDG & Pegasus \\
109 & 4177$\,\pm\,$16 & -1.05 & 54.0 & 4.99 & -15.28 & UDG & Pegasus
\enddata
\end{deluxetable*}

\begin{figure*}
    \centering
    \includegraphics[width=\textwidth]{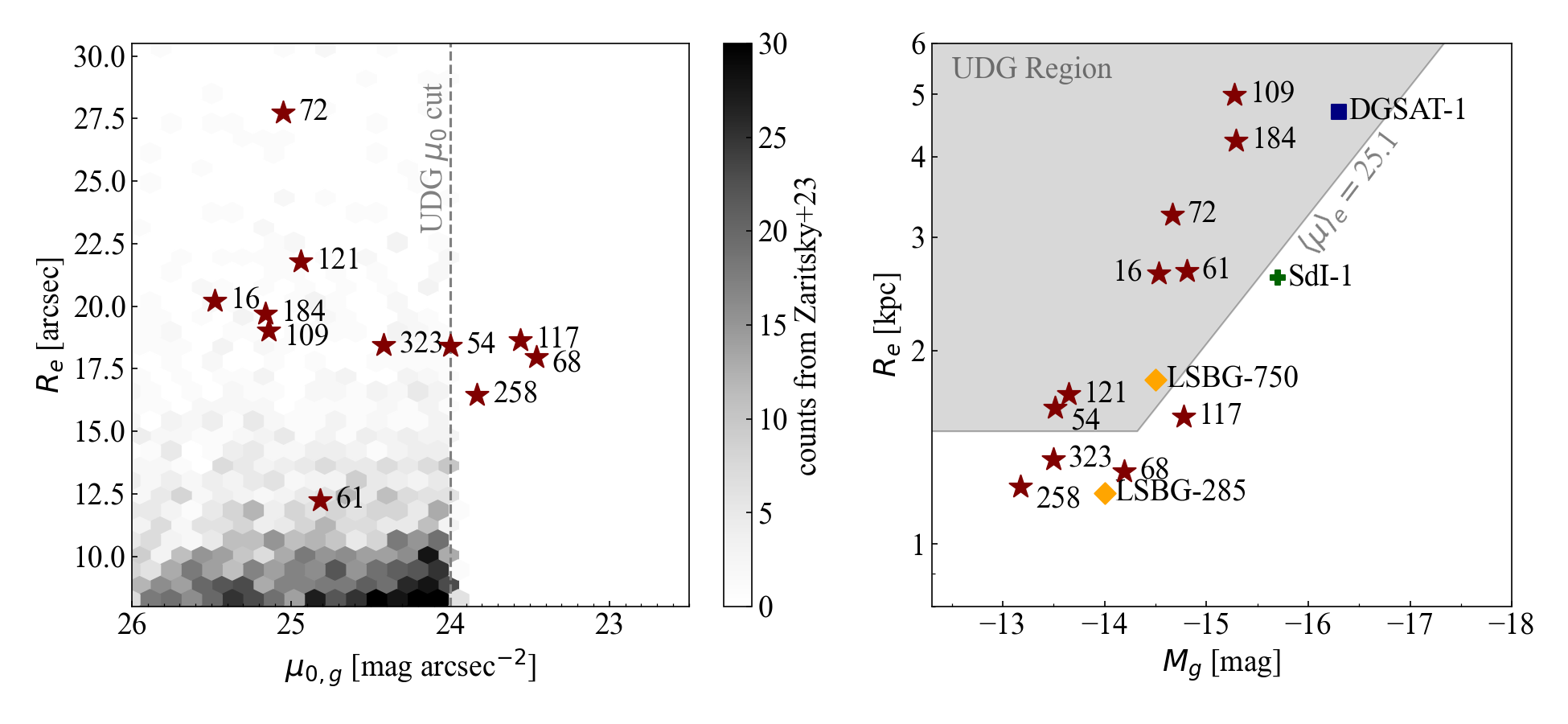}
    \caption{Left: the apparent effective radius $R_{\mathrm{e}}$ vs. the $g$-band central surface brightness $\mu_{0,g}$ of each galaxy in our sample (marked by red stars). The hexagons are a density map of UDG candidates from the SMUDGes survey \citep{Zaritsky2023ApJS} and the dotted vertical line represents the surface brightness definition of UDGs ($\mu_0(g) > 24\,\mathrm{mag\,arcsec^{-2}}$). Right: the physical size of each galaxy vs. the $g$-band absolute magnitude. Red stars represent galaxies in our sample, while the other colored symbols represent other quiescent isolated UDGs in literature \citep{Martinez-Delgado2016AJ,Bellazzini2017MNRAS,Greco2018ApJ}. The grey shaded region represents UDGs: the diagonal boundary shows constant ${\langle \mu\rangle}_e$ assuming $\mu_0(g) = 24$ and S{\'e}rsic $n=1$.}
    \label{fig:size-mass}
\end{figure*}

With the new distances, we can transform the observed quantities to physical properties. Table \ref{table:dist} lists the absolute magnitude and physical sizes of the galaxies. We classify each galaxy according to its $g$-band central surface brightness and physical effective radius. UDGs are defined as galaxies that satisfy $\mu_0(g) > 24\,\mathrm{mag\,arcsec^{-2}}$ and $R_e \geq 1.5$\, kpc \citep{vanDokkum2015ApJL}. LSB dwarfs are defined as galaxies that have $\mu_0(g) > 24\,\mathrm{mag\,arcsec^{-2}}$ and $R_e<1.5$\, kpc. The rest of the galaxies with $\mu_0(g) < 24\,\mathrm{mag\,arcsec^{-2}}$ are simply designated dwarf galaxies. In our sample, we find seven UDGs, one LSB dwarf, and three dwarf galaxies. 

To put these galaxies into context, Fig. \ref{fig:size-mass} is a summary of their apparent (left panel) and physical properties (right panel). The left panel shows their apparent size as a function of $g$-band central surface brightness. The background density plot is from the UDG candidates of SMUDGes \citep{Zaritsky2023ApJS}, where most are smaller than 10\arcsec. Our sample galaxies overlap with the biggest candidates in SMUDGes, and the cross-matched entries are in Table \ref{table:phot}. 
For the galaxies in common, our measured central surface brightness agrees well with the SMUDGes catalog but there is a $\sim3\arcsec$ difference in $R_e$.
The right panel of Fig \ref{fig:size-mass} shows our sample galaxies in the physical size -- absolute magnitude plane. Diagonal lines in this plot are of constant surface brightness, and UDGs occupy the upper left region. We also include other confirmed quenched UDGs in the field (yellow, green, and blue symbols); they are extremely rare.

Comparing with past spectroscopic campaigns on UDG candidates, we obtained a higher success rate and a similar UDG fraction. \citet{Kadowaki2021ApJ} carried out spectroscopic follow-up on 23 candidates from an earlier SMUDGes catalog and obtained radial velocities for 14 of them. Thanks to the light bucket strategy with KCWI, we were able to measure radial velocities for 100\% of the observed targets of similar surface brightness. We find that $7/11 = 64\%$ of the sample satisfied the physical size criteria for UDGs, which is very close to the UDG fraction of $\sim 60\%$ in \citet{Kadowaki2021ApJ}.

\subsection{Stellar populations} \label{sec:sps}
\begin{figure*}
    \centering
    \includegraphics[width=\textwidth]{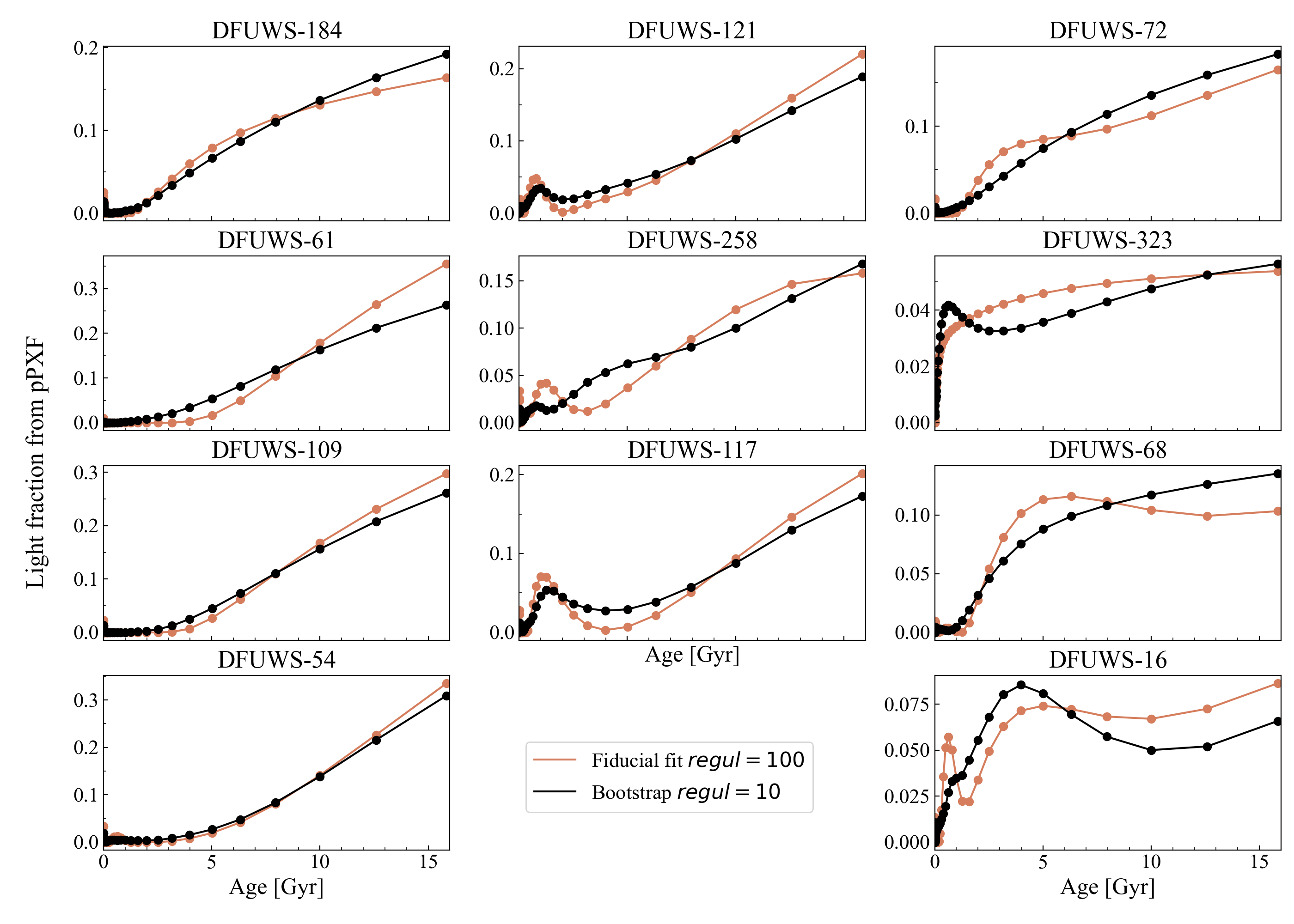}
    \caption{pPXF stellar population fits of KCWI spectra. The galaxies in the left column require one star formation event, while the galaxies in the middle and right columns require multiple star formation events. For each galaxy, the orange line shows the light distribution as a function of template age in a single fit with a relatively high regularization parameter. The black line shows the average distribution from 30 bootstrap resampling trials using a lower regularization parameter. }
    \label{fig:ppxf-sps}
\end{figure*}
The latest version of \texttt{pPXF} allows for simultaneously deriving the radial velocity and the stellar populations that best fit the data. \texttt{pPXF} uses a non-parametric grid of models and finds the optimal distribution of weights across this grid. Following \citet{Cappellari2023MNRAS}, we report the fraction of bolometric luminosity contributed by each template instead of the stellar mass.
We chose a relatively high regularization (\texttt{regul=100}) for our fiducial fit because this worked well in injection-recovery tests that simulate our data quality (see Appendix \ref{appendix}). The best fit model is shown in Fig. \ref{fig:spec} as the orange line, and the weights distribution in age is shown as the black points in Fig. \ref{fig:ppxf-sps}. Regularization is a trade-off between agreement with the data and smoothness in the weights distribution. Higher regularization reduces noise and spurious features in the solution, but it could overly smooth the underlying weight distribution. We also performed 30 rounds of bootstrap resampling using low regularization and averaged the output weights.
We implemented bootstrap resampling by randomly adding back the residuals from our fiducial fit to the fiducial model, then re-running the fit routine, keeping all parameters except the regularization the same as the fiducial fit. The lower regularization (\texttt{regul=10}) leads to a more discontinuous solution for each bootstrap run, but averaging over many runs gives a smooth distribution of weights again. This averaged bootstrapped distribution for each galaxy is shown as the orange line in Fig. \ref{fig:ppxf-sps}. 

Given the spectral resolution of our data, we do not aim to reconstruct a detailed star formation history; instead, we investigate whether our data are compatible with a single stellar population. 
The left column in Fig. \ref{fig:ppxf-sps} shows galaxies that can only be described by a single star formation event at a localized age, regardless of the value of the \texttt{regul} parameter. All three galaxies that are in cluster environment (184, 61, and 109, see Section \ref{subsec:env-cluster}) are in this category, consistent with a single old stellar population.
Galaxies shown in the middle column of Fig. \ref{fig:ppxf-sps}) need a $\sim 1$\,Gyr population in addition to an old metal-poor population to reproduce the data. There is good agreement between the single fit with \texttt{regul=100} and the bootstrapped fits with \texttt{regul=10}.
For galaxies in the right column of Fig. \ref{fig:ppxf-sps}, the two methods produce different age distributions likely due to the different choice in parameterization, but both models require extended star formation events and neither model is consistent with a single old metal-poor stellar population.

In the injection-recovery tests and in our fitting process, we found that the main spectral features that require multiple populations are the prominent Balmer lines and the ratio of the Ca II H line to the Ca II K line. We note that these features could also be explained by a prominent blue horizontal branch \citep{Schiavon2004ApJL,Cabrera-Ziri2022MNRAS}, which we discuss further in Section \ref{subsec:bhb}.

\subsection{Environment}
The environment plays a crucial role in understanding the evolution of galaxies. Factors such as tidal forces and gas stripping can significantly impact the stellar populations and overall properties of galaxies in high density environments such as clusters.
The galaxies in this sample span a range of environments: three are in the Pegasus cluster, five are in groups, and the rest are in low density environments.

\subsubsection{Cluster} \label{subsec:env-cluster} Galaxies 184, 61, and 109 are located in the outskirts of the Pegasus I cluster. The Pegasus cluster has an average velocity of $3885 \pm 32 \,\mathrm{km\,s^{-1}}$  and a velocity dispersion of $236\pm 33\,\mathrm{km\,s^{-1}}$ from 75 confirmed members \citep{Richter1982A&A}. The distance to the group is 51.8 Mpc. Previous photometric surveys in the Pegasus cluster \citep{Shi2017ApJ, Gonzalez2018A&A} have found numerous UDGs in the cluster but they did not obtain spectroscopic follow-up. While we will quantify the completeness in future work, other catalogued UDGs are detected in DFUWS from our visual inspection. The projected separations of all three galaxies from the central galaxy NGC 7619 are respectively 51, 28, and 48 arcmin, all within the Abell radius of Pegasus I \citep[$2.36\arcdeg$, ][]{Solanes2001ApJ}.
The velocity separations from NGC 7619 are respectively $-447$, $-410$, and $178 \,\mathrm{km\,s^{-1}}$.

\subsubsection{Group} 
We searched the \citet{Kourkchi2017ApJ} group catalogue to find possible group associations for each galaxy in our sample. We use the projected Second Turnaround Radius $R_{2t}$ as the size of the group, the heliocentric radial velocity of the group $v_h$, and the line-of-sight velocity dispersion of the group $\sigma_v$ based on the radial velocities of its members. For each galaxy in our sample, we search for any group that is within $1.15 \times R_{2t}$ of our galaxy and satisfies the velocity criteria $ | v - v_h | < 2 \sigma_v $. 
The group distances in the catalogue had large uncertainties and our distance was based on the velocity, so we did not include a distance criteria in the search.
We identified five galaxies that are part of a group. 

\paragraph{DFUWS-68} Galaxy 68 is in the NGC\,7814 group with a relative velocity of $70\,\mathrm{km\,s^{-1}} $.
This group has a velocity dispersion of $\sigma_v = 92\,\mathrm{km\,s^{-1}}$ and a group radius is $R_{2t}$ = 315\,kpc = $110\arcmin$.
%Group distance is 9.84Mpc
This group has seven previously identified dwarf galaxies from the DGSAT survey \citep{Javanmardi2016A&A, Henkel2017A&A}. We detect all seven known dwarf galaxies in DFUWS and add one more.
At 460 kpc projected separation, DFUWS-68 is the dwarf galaxy that is the farthest from the center of the group.

\paragraph{DFUWS-117} Galaxy 117 is 19$\arcmin$ away from the NGC\,7332/NGC\,7339 group.
%The group distance is 22.55 Mpc, and 
The group radius is $R_{2t}  = 338\,\mathrm{kpc} = 51.5 \arcmin$.
The relative velocity between DFUWS-117 and the group is $-21\,\mathrm{km\,s^{-1}}$, within the group velocity dispersion of $\sigma_v = 91\,\mathrm{km\,s^{-1}}$.
The central region of this group had been imaged by the DGSAT survey \citep{Henkel2017A&A} but their field of view did not cover DFUWS-117.

\paragraph{DFUWS-54} Galaxy 54 is in the NGC\,1052 group which is at 20 Mpc. It is within the group virial radius  $R_{2t} = 528\,\mathrm{kpc} = 106 \arcmin$. 
Galaxy 54 has a radial velocity of $-20\,\mathrm{km\,s^{-1}}$ with respect to NGC\,1052 and it is within the velocity dispersion $\sigma_v = 143\,\mathrm{km\,s^{-1}}$. This group has been extensively searched for LSB galaxies, but the existing catalog from \citet{Roman2021A&A} likely did not include DFUWS-54 due to its central star cluster. 

\paragraph{DFUWS-72} Galaxy 72 is $21.51\arcmin$ from the center of the NGC\,7454 group.
At a distance of 26 Mpc, this group has a radius of $R_{2t} = 418 \,\mathrm{kpc} = 55 \arcmin$.
The radial velocity of DFUWS-72 with respect to the group is $-139\,\mathrm{km\,s^{-1}}$ which is within the group velocity dispersion $\sigma_v = 143\,\mathrm{km\,s^{-1}}$.
DFUWS-72 was first identified in the MATLAS survey \citep{Poulain2021MNRAS} which focused on group environments.

\paragraph{DFUWS-16} Galaxy 16 is closest to the NGC\,488 group with an on-sky separation of $50\arcmin$.
The NGC\,488 group is at a distance of 33\,Mpc, and has a radius of $R_{2t} = 665\,\mathrm{kpc} = 70 \arcmin$.
The radial velocity of DFUWS-16 with respect to NGC\,488 is $-50\, \mathrm{km\,s^{-1}}$, within the group velocity dispersion is $\sigma_v = 80\,\mathrm{km\,s^{-1}}$.  

\subsubsection{Field}

To investigate the environments of the galaxies which are not in groups and to catch any potential missed group associations, we queried the NASA-Sloan Atlas \footnote{http://nsatlas.org/} (NSA) catalog.
All of the galaxies in our sample are well within the NSA footprint, which has the coverage of SDSS DR8 \citep{Aihara2011ApJS}. 
The NSA provides a consistent catalogue of galaxies with known redshifts out to $z=0.055$ and stellar mass estimates based on SDSS and \textit{GALEX} fluxes where available. 

We estimated the virial radius of each NSA galaxy from its catalogued stellar mass. Then we queried the NSA catalogue with very generous criteria: we included any NSA galaxy above $M_*>10^8 M_{\odot}$ if there was a DFUWS galaxy within $3\times R_{vir}$ in the projected on-sky separation and within a relative velocity $\delta v < 500\, \mathrm{km\,s^{-1}}$. Similar to the last search, we did not use the distance information for DFUWS galaxies in the query.
After finding a match, we record its NSA ID number and match it back to the \citet{Kourkchi2017ApJ} catalogue to find the group ID. 

258 and 121 are at similar distance and velocity. The group that they are closest to is NGC 936, at a projected 580 kpc and 510 kpc respectively. The Second turnaround radius of the group is 447 kpc, which makes it unlikely that either galaxy is associated with the group.

The closest massive galaxy to 258 is UGC\,1862, at a projected separation of 270\,kpc and a relative velocity of $55 \, \mathrm{km\,s^{-1}}$.

DFUWS 323 is closest to NGC 428 group with a relative velocity of $90\, \mathrm{km\,s^{-1}}$, but the groups Second turnaround radius is 140kpc \citep{Kourkchi2017ApJ} and DFUWS 323 is at a projected distance of 320 kpc.

Using the criteria of \citet{Kadowaki2021ApJ} (not within 300\,kpc from any massive galaxy with a relative velocity difference $< 500 \mathrm{km\,s^{-1}}$), 121 and 323 are in sparse environments.

\section{Discussion and Conclusion} \label{sec:dis}
In this paper, we presented the discovery and analysis of galaxies found using early data from the DFUWS. The area around the South Galactic Cap was visually inspected for LSB galaxies, and this sample contains galaxies with the biggest apparent sizes in these 3100 deg$^2$. We obtained spectroscopic followup with KCWI and determined their radial velocities, distances, and physical properties, enabling us to classify these galaxies as UDGs, LSB dwarfs, and dwarf galaxies. The location of each galaxy in clusters, groups, or low-density environments is an important piece in understanding the evolution of these galaxies. We also find evidence of multiple stellar populations in this sample.
\subsection{The nature of the largest LSB galaxies}
From our visual search of 3100 deg$^2$ Dragonfly data, we have not identified any giant galaxies that exceed an arcminute in radius. The largest LSB galaxies in our sample have mostly been identified by existing surveys and are all detected in Legacy data. If there existed giant LSB galaxies that are not seen in previous imaging, we would have found them in the DFUWS data. Although we cannot quantify the completeness of the process, Dragonfly has demonstrated high sensitivity for large scale LSB emission (e.g. in cirrus and tidal features in the outskirts of galaxies) and our visual search is not limited by a pre-defined set of object detection criteria. This implies that arcminute-sized galaxies are exceedingly rare, or fainter than $\sim27$ mag, or coincide with cirrus-rich regions, if they exist at all.
We also find that most of these large LSB galaxies are UDGs at around 20\,Mpc instead of nearby dwarf galaxies. Numerous dwarfs were predicted around the Milky Way as ``stealth galaxies'' due to their low surface brightness \citep{Bullock2010ApJ}. In this sample, we do not find any dwarf galaxies within 10\,Mpc despite our sensitivity.
We aim to quantify the completeness when we have an automated detection pipeline, but so far we have not found any truly unexpected galaxies.

\subsection{The origin of quiescent UDGs in low density regions}
The stellar populations of galaxies in our sample show a remarkable dependence on the environment. All three UDGs in the Pegasus cluster do not show strong Balmer lines and are consistent with an old stellar population quenched 10\,Gyr ago by the cluster. In the rest of the sample, seven out of eight galaxies need multiple stellar populations to reconstruct their spectra. 
This result is consistent with previous KCWI spectra of 14 UDGs \citep{Ferre-Mateu2023MNRAS}, which showed that UDGs in high-density environments tend to form earlier and faster than those in low-density environments. 

While the picture is clear for cluster UDGs, the formation of UDGs in low-density environments is more puzzling.
The extended star formation is inconsistent with the ``failed Milky Way'' scenario \citep{vanDokkum2017ApJL} where UDGs occupy massive halos but were quenched before forming a disk and bulge, and the backsplash orbit scenario \citep{Benavides2021NatAs} where the field UDGs were quenched by interaction with a cluster.
The natural hypothesis is that the large sizes and the multiple star formation events of these galaxies share a common origin. For example, the galaxies could be puffed up by stellar feedback from globular clusters (GCs) that formed during the initial collapse \citep{Trujillo-Gomez2022MNRAS}. This scenario requires massive GC systems to drive unusually strong feedback, so the GCs of these field UDGs may provide additional clues to their formation.

\subsection{Stellar population alternatives} \label{subsec:bhb}
It is clear from the spectra that the dominant stellar population in our sample is old and metal-poor. The red colors and prominent Balmer lines in the sample are consistent with recently quenched galaxies. The apparent presence of an intermediate-age population, however, is not the only possible explanation for the observed Balmer lines. \citet{Kadowaki2017ApJL} found Balmer lines in Coma cluster UDGs whose spectra resemble those of A-stars. They compared the UDG spectra to old SPS templates with a range of metallicities as well as a 1\,Gyr template by eye. They interpreted the spectra as an old, metal-poor stellar population, but did not discuss the possibility of multiple ages. Our injection-recovery tests with two-component spectra (see Appendix \ref{appendix}) show that \texttt{pPXF} can quantify the contribution of each.

The other possible reason for prominent Balmer lines is horizontal branch (HB) stars.
It is well-known that ages derived from the analysis of integrated galaxy light can be significantly biased towards younger values by the presence of hot HB stars \citep{Schiavon2004ApJL, Conroy2018ApJ}. The SPS models that were used in our full-spectrum fitting routine assumes a standard set of parameters which do not account for possible variation in HB morphology.
The luminosity, temperatures, and spectral lines (e.g. Balmer lines) of extended HB stars are similar to young main-sequence stars, 
so if a galaxy has extended HB but is modeled with standard SPS templates, a younger age is often required to reproduce its spectrum.
This problem has been extensively studied for globular clusters \citep{Schiavon2004ApJL,Cabrera-Ziri2022MNRAS}, and an extended HB component can be added to the model if we have \textit{a priori} knowledge that the stars should all have the same age.
Without this assumption, integrated light spectra of galaxies do not contain enough information to distinguish between a young stellar population and an extended HB.
While we cannot rule out an extended HB based on the spectra, there is additional evidence from the environment of these galaxies. The observed environmental dependence (single population in clusters, multiple populations outside of a cluster) would be harder to explain if the Balmer lines were due to an extended HB. 

To summarize, the \texttt{pPXF} results indicate that seven galaxies in our sample have multiple stellar populations or extended star formation events. 
The complex stellar populations of these galaxies could be connected to their diffuse nature. These topics will be explored in a future paper.

\subsection{Full survey}
\label{subsec:mrf}
This paper presents a pilot sample of galaxies that we identified from DFUWS early data.
The next step will be to identify UDG candidates from the full survey using an automated detection method.
Since Oct 2023, there have been a number of improvements in the data reduction pipeline and the survey coverage has increased. 
We have used the pilot sample to develop a detection method for LSB galaxies in Dragonfly data based on the existing Multi-Resolution Filtering method \citep[MRF,][]{vanDokkum2020PASP}. MRF subtracts high surface brightness sources from Dragonfly by comparing with a higher resolution image. 
%This process is illustrated in Fig. \ref{fig:legacy-uw-comparison}. The left panel shows a cutout from DFUWS data that contains a UDG among other stars and galaxies. The middle panel is the Legacy image at the corresponding location, which functions as the high-res data. Compact sources are detected in the high-res image, convolved with the Dragonfly PSF, and then subtracted from the original Dragonfly image. The result (right panel) shows a clean image with only the LSB galaxy left behind. Notably, MRF can preserve the UDG without any prior knowledge of its location. It simply subtracts everything that is compact and/or high surface brightness. We used the eleven sample galaxies in this paper to tune the MRF parameters until everything other than the UDGs are subtracted. This method would also work well with anything not detected in Legacy (e.g. tidal features, galactic cirrus). 
We plan to run MRF across the entire DFUWS footprint with a uniform set of parameters determined from this pilot sample. This will be the main step in the detection pipeline of the full survey.
To estimate the completeness of the full catalog, we also plan to inject galaxies using \texttt{ArtPop} \citep{Greco2022ApJ} and find the recovery fraction. 

In future work, we plan to release a full catalogue of the UDG candidates from the entire DFUWS footprint.
As with this work, we will focus on the galaxies with the largest apparent size and obtain spectroscopic follow-up to measure their distances. We have established a method that reliably yields adequate signal-to-noise for LSB galaxies, and we aim to obtain spectra for as many candidates from the full survey as possible.

\begin{acknowledgments}
We thank Charlie Conroy for comments on the stellar population analysis. 
Chloe Neufeld and Josephine Baggen contributed to data inspection. We also thank Rosalie McGurk for her help with the KCWI observations.

\end{acknowledgments}

\vspace{5mm}
\facilities{Keck(KCWI)}

\software{astropy \citep{astropy:2013,astropy:2018,astropy:2022},  
          AutoPhot \citep{Stone2023MNRAS}, 
          pPXF \citep{Cappellari2023MNRAS}
          }

\appendix
\section{Injection Recovery test for \texttt{pPXF}} \label{appendix}
\begin{figure*}[h!]
    \plottwo{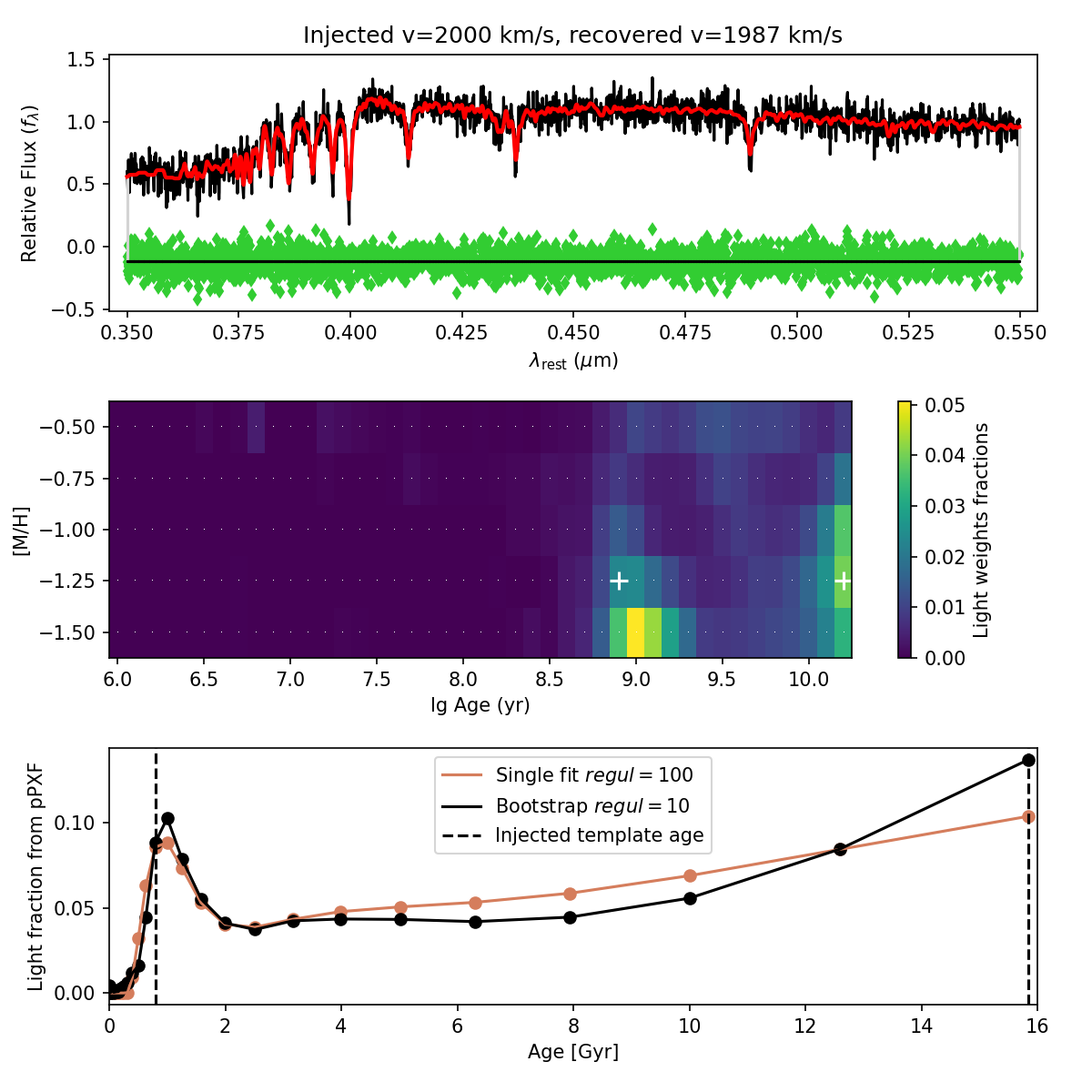}{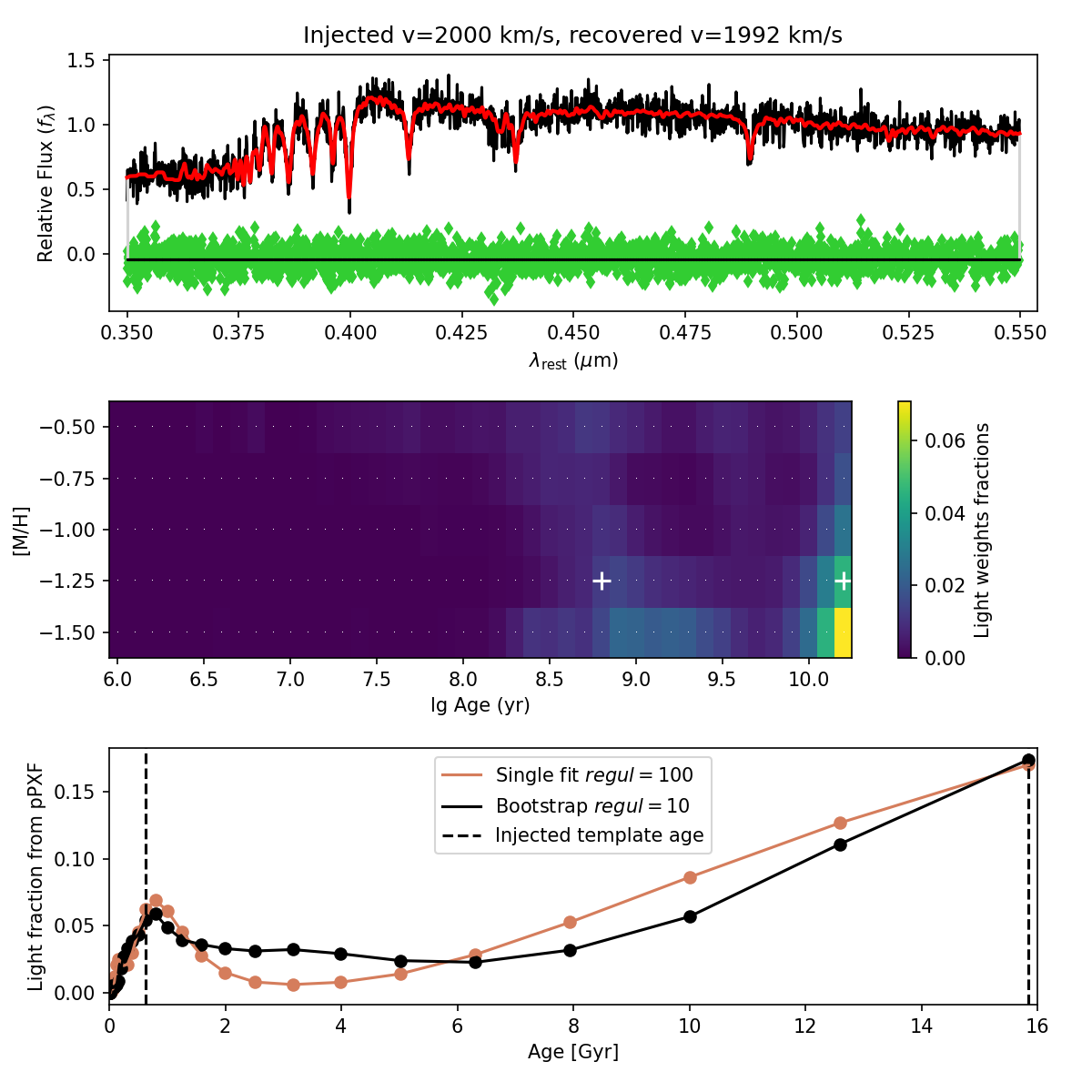}
    \caption{Two examples of injection-recovery tests conducted with \texttt{pPXF}, where the only difference is a 0.1 dex change in the age of the injected templates (the white crosses in each panel). Top panels: a mock galaxy spectrum (black) with noise added (seen in the residuals in green diamonds) to approximate the signal-to-noise ratio of the data. The \texttt{pPXF} best fit is overplotted in red. The mock galaxy spectrum is a linear combination of two templates, whose age and metallicity are shown in the middle panel as white crosses. Middle panels: The weight of each template that went into the \texttt{pPXF} best fit. Bottom panels: the weights distribution summed along the metallicity axis, plotted against linear age. The peak at $\sim1$ Gyr and at old ages is clearly visible. The input spectrum on the left and right are very similar, but the reconstructed metallicity distributions are quite different. We conclude that at the signal-to-noise level of our data, only the age distribution can be reliably recovered. }
    \label{fig:injection-test}
\end{figure*}
We performed a series of mock data tests to see how well  \texttt{pPXF} can recover different star formation histories (SFHs) given our data quality.
We started by choosing two single stellar population templates from the same SSP library that was used in the fiducial fit and combining them with different weights. We broadened the mock spectrum to $1.1$\AA, or $140 \, \mathrm{km\,s^{-1}}$, which is close to the instrumental resolution measured from the data. We redshifted the mock spectra to have radial velocity $v= 2000 \, \mathrm{km\,s^{-1}}$, and resample the mock spectrum to have the same pixel scale as the observed spectra. Finally, we added Gaussian noise with $\sigma=0.1$ which gives an average signal-to-noise ratio (SNR) of 10, close to the data. We fit the mock spectra with \texttt{pPXF} using the same SSP library for the fiducial fit and keep all other fitting parameters unchanged. Then we change the two inputs and repeat this process for all combinations of one old and one young template at a given metallicity. One of the mock spectra and the fit is shown in Fig. \ref{fig:injection-test}. The top panel shows the mock spectrum in black and the best-fit model in red. The spectra looks similar as real data shown in Fig. \ref{fig:spec}. The middle panel shows the best-fit weights for templates of different ages and metallicities. The weight distribution clearly peaks at the age of the two injected templates (marked by white crosses), but the metallicity is not accurately recovered. The weights are then summed along the metallicity axis and plotted against linear age in the bottom panel, where the true ages are indicated by vertical dashed lines. In most cases, we find good recovery of the input velocity and age, but not the metallicity.  We note that in injections tests with SNR $\sim20$, the underlying stellar populations are almost always recovered perfectly, so the spread in metallicity is likely due to the relatively low SNR of our data. Thus in the main body of the paper we choose to only present the age distribution. In addition, the input weights are the same between the left and right columns of Fig. \ref{fig:spec}, but due to the noise, the value of the weights assigned to the template changes.

\citet{Cappellari2023MNRAS} explained that a regularization scheme is needed because there is no unique solution to match templates to data. \citet{Kacharov2018MNRAS} performed injection tests on higher resolution and higher signal-to-noise simulations, and they caution that the regularization parameter imposes a prior for the model weights and the SFH is only perfectly recovered if it matches the prior. 
However, for the real data the SFH is unknown, so we present the results from both a high value and a low value of regularization. Regardless of the regularization choice, the age distribution is clearly inconsistent with a single population, and that is the main point we make for the data shown in this paper.

%% For this sample we use BibTeX plus aasjournals.bst to generate the
%% the bibliography. The sample631.bib file was populated from ADS. To
%% get the citations to show in the compiled file do the following:
%%
%% pdflatex sample631.tex
%% bibtext sample631
%% pdflatex sample631.tex
%% pdflatex sample631.tex

\bibliography{sample631}{}
\bibliographystyle{aasjournal}

\end{document}